\documentclass[conference]{IEEEtran}
\usepackage{color}
\usepackage{balance}
\usepackage{lipsum}

\usepackage{cite}

\ifCLASSINFOpdf
   \usepackage[pdftex]{graphicx}
\else
\fi
\usepackage{amsmath}
\usepackage{mathtools}
\usepackage{amssymb}
\usepackage{amsthm}
\usepackage[binary-units=true,per-mode=symbol,alsoload=binary]{siunitx}

\newcommand{\paren}[1]{\left(#1\right)}
\newcommand{\sqparen}[1]{\left[#1\right]}
\newcommand{\brparen}[1]{\left\{#1\right\}}

\newcommand{\ES}[1]{\ensuremath{\mathsf{E}\left[#1 \right]}} 

\usepackage{bm}

\usepackage{algorithmic}
\ifCLASSOPTIONcompsoc
    \usepackage[caption=false,font=normalsize,labelfont=sf,textfont=sf]{subfig}
\else
    \usepackage[caption=false,font=footnotesize]{subfig}
\fi

\begin{document}
\bstctlcite{BSTcontrol}
%
\title{Deep Reinforcement Learning-Based Topology Optimization for Self-Organized Wireless Sensor Networks}




\author{
    \IEEEauthorblockN{Xiangyue Meng\IEEEauthorrefmark{1}, Hazer Inaltekin\IEEEauthorrefmark{2}, and Brian Krongold\IEEEauthorrefmark{1}}
    \IEEEauthorblockA{\IEEEauthorrefmark{1}Department of Electrical and Electronic Engineering, The University of Melbourne, Melbourne, VIC 3010, Australia.
    \\\{xiangyuem@student., bsk@\}unimelb.edu.au}
    \IEEEauthorblockA{\IEEEauthorrefmark{2} School of Engineering, Macquarie University, North Ryde, NSW 2109, Australia.
    \\ hazer.inaltekin@mq.edu.au}
}



\maketitle

\begin{abstract}
Wireless sensor networks (WSNs) are the foundation of the Internet of Things (IoT), and in the era of the fifth generation of wireless communication networks, they are envisioned to be truly ubiquitous, reliable, scalable, and energy efficient. To this end, topology control is an important mechanism to realize self-organized WSNs that are capable of adapting to the dynamics of the environment. Topology optimization is combinatorial in nature, and generally is NP-hard to solve. Most existing algorithms leverage heuristic rules to reduce the number of search candidates so as to obtain a suboptimal solution in a certain sense. In this paper, we propose a deep reinforcement learning-based topology optimization algorithm, a unified search framework, for self-organized energy-efficient WSNs. Specifically, the proposed algorithm uses a deep neural network to guide a Monte Carlo tree search to roll out simulations, and the results from the tree search reinforce the learning of the neural network. In addition, the proposed algorithm is an anytime algorithm that keeps improving the solution with an increasing amount of computing resources. Various simulations show that the proposed algorithm achieves better performance as compared to heuristic solutions, and is capable of adapting to environment and network changes without restarting the algorithm from scratch.
\end{abstract}


%
\IEEEpeerreviewmaketitle

\section{Introduction}
The Internet of Things (IoT) has emerged as a new communication paradigm where a huge number of heterogeneous physical sensing devices are seamlessly interconnected to autonomously collect information without human aid. Being the foundation of IoT, wireless sensor networks (WSNs) collect sensing data and forward the data to the core network for further processing. With the advent of the fifth generation (5G) of wireless communication networks, WSNs are envisioned to be truly ubiquitous, reliable, scalable, and energy-efficient \cite{Palattella2016}. To this end, a framework of Self-Organized Things was first introduced in \cite{Akgul2016}, where the sensors undergo automatic configurations to maintain connectivity and coverage, reduce energy consumption, and prolong network lifetime.

In a typical WSN, wireless sensors continuously monitor the environment and periodically generate small amounts of data. The data needs to be forwarded to another sensor for data aggregation or directly transmitted to the gateway. Since the sensors are generally battery-powered, energy efficiency is a prominent need to prolong the lifetime of the network, especially for low-power wide-area networks \cite{Kaur2017}. The major part of the energy stored in a sensor is consumed during data transmission, and the  energy consumption increases exponentially with the transmission distance \cite{Martinez2015}. Therefore, a multi-hop tree topology for aggregating sensor data at a gateway, at which the tree is rooted, has the advantage of reducing the per-sensor energy consumption, especially for sensors at the edge of the WSN, by decreasing transmission distances. In addition, tree topologies eliminate the cost of maintaining a routing table at each sensor, when compared to mesh topologies.

\begin{figure}[!t]
    \centering
    \includegraphics[width=0.48\textwidth]{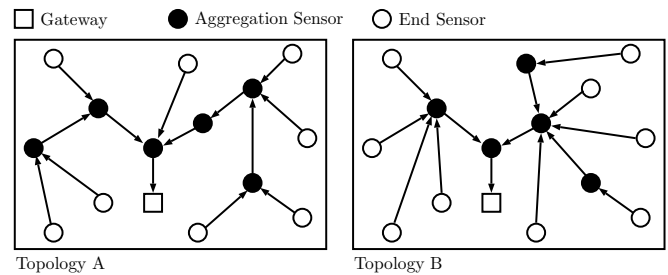}
    \caption{Two possible topologies of a WSN, rooted at the gateway.}
    \label{fig:rtree}
    \vspace{-1em}
\end{figure}

Figure \ref{fig:rtree} shows two possible topologies of a WSN rooted at a gateway. Finding the optimal topology in terms of energy efficiency in a WSN is combinatorial in nature and NP-hard to solve \cite{Wu2008}. The exhaustive search is not practical because the number of connected devices in an IoT system today is usually very large and achieving the optimal network configuration via exhaustive search is exponentially complex due to the tremendously large search space of all possible topology configurations. Most of the existing work in the literature leverage properties of a specific network model to heuristically reduce the number of potential search candidates. However, with the growing heterogeneity of WSNs in the 5G era, a unified topology optimization framework is desirable so as to seamlessly utilize various IoT technologies and adapt to the dynamics of the environment.

In this paper, inspired by the success of deep learning achieving human-level proficiency in many challenging domains, we propose a deep reinforcement learning-based topology control (DRL-TC) algorithm as a generic approach to optimize the network topology for energy-efficient WSNs in the face of heterogeneity and uncertainties in the networks, without relying on any domain knowledge beyond the topology rules. To be specific, the proposed DRL-TC employs the framework of deep reinforcement learning (DRL) with a Monte Carlo tree search (MCTS) to sequentially construct the network according to pre-defined topology rules. A deep neural network (DNN) is trained to predict the energy consumption of a partially-built topology and guides the MCTS to roll out the remaining steps in more promising areas in the search space. In return, the search results from the MCTS reinforce the learning of the DNN to obtain a more accurate prediction in the next iteration. Our contributions are as follows:
\begin{itemize}
    \item We propose a novel and generic DRL-TC algorithm to determine a near-optimal topology for WSNs in terms of energy efficiency without relying on specific domain knowledge beyond topology rules.
    \item The proposed algorithm is a statistical anytime algorithm\footnote{An anytime algorithm continuously returns valid solutions before it ends. It can also resume at anytime without restarting from scratch.} that is capable of adapting to the dynamics of the environment (including possible unexpected network changes) and re-configures the network accordingly.
    \item Various simulation results show that the proposed DRL-TC outperforms other heuristic approaches to a large extent.
\end{itemize}

\section{Related Work}
Different from cellular networks, the networks of IoT devices generate small amounts of data and are expected to be operational over long time periods with limited battery powers. Hence, instead of maximizing the network throughput, a prominent objective of an IoT-WSN is to minimize energy consumption in order to maximize the network lifetime, subject to the constraints of coverage and reliability \cite{Kaur2017}.

In the literature, LEACH in \cite{Heinzelman2002} and its many variants belong to the class of distributed and cluster-based algorithms where a local cluster of sensors elects a cluster head at a time to aggregate the data and forwards the aggregated data to the gateway. LEACH periodically rotates the role of cluster head depending on the residual energy of the sensors in each cluster. It has the advantages of scalability and ease of implementation, but requires the sensors constantly exchanging information with each other, which introduce an extra amount of energy consumption at each sensor. Similarly, the authors in \cite{Xu2016a} proposed a joint clustering and routing algorithm for sensor data collection. The authors in \cite{Wang2018d} considered the case of non-uniform traffic distribution for load balancing and energy efficiency.

On the other hand, centralized algorithms relieve the burden of end sensor computations, but generally have high computing complexity. In order to reduce this complexity while maximizing a WSN's lifetime, the authors in \cite{Imon2015} proposed a load balancing approach which randomly switches some sensors from their original paths to other paths with a lower load. The authors in \cite{Zhu2019} proposed a tree-based algorithm with a set of heuristic rules to construct a tree topology in multi-hop WSNs. In \cite{Naznin2015}, the authors reduced the search space of tree topologies by dividing the network into on-demand data collection zones and routes.

With the fast development of the theory and practice of deep learning, deep reinforcement learning (DRL) has become a powerful paradigm in many areas of wireless communications, such as network optimization, resource allocation, and radio control \cite{Zhang2019a}. Integrating DRL with MCTS, AlphaGo Zero from Google demonstrated exceedingly superhuman proficiency in playing the game of Go \cite{Silver2017a}. DRL-MCTS is a very powerful framework for solving the problems where sequential decisions are required to achieve a final outcome, which usually results in an NP-hard problem.

The study of DRL-MCTS in the context of wireless communications is very limited. In this paper, we employ DRL-MCTS for the topology optimization in WSNs and propose a deep reinforcement learning-based topology control (DRL-TC) algorithm to sequentially construct the topology of a WSN with the objective of minimizing the energy consumption at each sensor.

\section{System Model and Problem Formulation}

\begin{figure}[!t]
    \centering
    \includegraphics[width=0.49\textwidth]{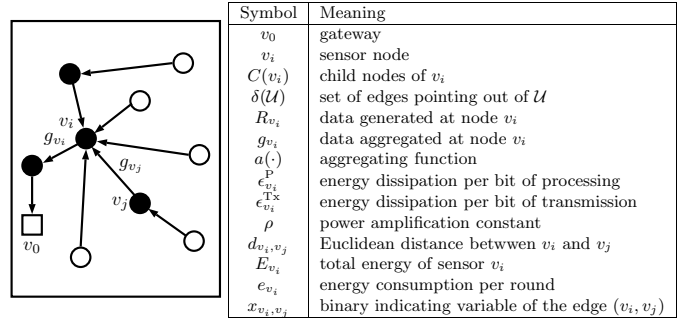}
    \caption{Notations used in the network model.}
    \label{fig:model}
    \vspace{-1em}
\end{figure}

\subsection{System Model}
We consider the uplink of a WSN consisting of IoT devices that collect raw data and forward the collected data to the core network. As shown in Fig. \ref{fig:model}, the WSN has a single gateway $v_0$, and $N-1$ sensors $\brparen{v_1,v_2,\ldots,v_{N-1}}$. Denoting $V=\brparen{v_0,v_1,\ldots,v_{N-1}}$ as the set of all vertices, and $E$ as the set of directed edges, we model the WSN as an \textit{arborescence}\footnote{An arborescence is a directed, rooted tree $T=(V,E)$ that spans (when viewed as an undirected graph) the graph.} where every sensor has a unique path to the gateway $v_0$.

In each round of data collection, the sensor $v_i,i\in\brparen{1,2,\ldots,N-1}$, needs to forward
\begin{equation}
    g_{v_i}=R_{v_i}+a\paren{\textstyle{\sum}_{v_j\in C(i)}R_{v_j}}
\end{equation}
bits of data to its parent sensor, where $v_i$ generates $R_{v_i}$ bits of its own data and aggregate the data $\sum_{v_j}R_{v_j}$ from its child sensors $v_j\in C(i)$, and $a(\cdot)$ is an aggregating function. We adopt the energy consumption model in \cite{Martinez2015}, where the topology-relevant energy consumption at sensor $v_i$ largely consists of data processing (including data receiving) and transmitting energy consumption. This is modeled as
\begin{equation}
    e_{v_i}=\paren{\epsilon^\text{P}_{v_i}+\epsilon^\text{Tx}_{v_i}}g_{v_i},
\end{equation}
where $\epsilon^\text{P}_{v_i}$ and $\epsilon^\text{Tx}_{v_i}$ is the energy dissipation per bit for data processing and transmission at sensor $v_i$, respectively. The energy dissipation per bit for data transmission depends on the distance to the parent sensor, and is further modeled as
\begin{equation}
    \epsilon^\text{Tx}_{v_i}=\rho d_{v_{i},v_{j}}^2,
\end{equation}
where $d_{v_{i},v_{j}}$ is the Euclidean distance between $v_i$ and its parent sensor (or the gateway) $v_j$, and $\rho$ is a constant of power amplification in the link budget, considering the effects of shadowing and fading.

\subsection{Problem Formulation}
The proposed energy-efficient topology optimization framework follows the general setting below:
\begin{enumerate}
    \item The data size $R_{v_i}$ generated by sensor $v_i$ is a random number drawn from a certain distribution that is unknown to the DRL-TC algorithm.
    \item The aggregating function $a(\cdot)$ can be any deterministic function. For the purpose of demonstration, we use summation in this paper.
    \item
    The designed topology control algorithm should be readily applicable to other network objectives, such as minimizing the overall network energy consumption or maximizing the network throughput.
\end{enumerate}

\begin{figure}[!t]
    \centering
    \includegraphics[width=0.49\textwidth]{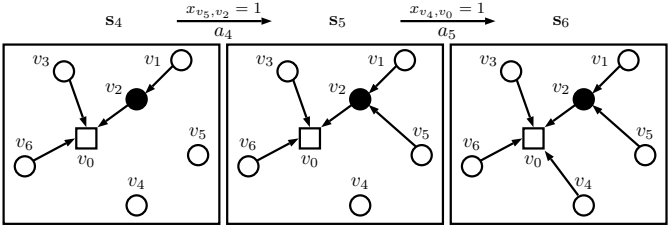}
    \caption{The MDP of constructing an arborescence in a WSN with $7$ sensors starting at step $4$ and completed after two steps.}
    \label{fig:mdp}
    \vspace{-1em}
\end{figure}

Denote the total battery energy stored at sensor $v_i$ as $E_{v_i}$, and let $E_{v_0}=\infty$ since the gateway $v_0$ is assumed to be connected to an unlimited main power supply. We define the lifetime of the WSN as the minimum battery lifetime of all sensors in terms of the total rounds of transmission. This lifetime maximization of the WSN can be formulated as
\begin{IEEEeqnarray}{rll}\label{eqn:opt}
    \underset{\{x_{ij}\}}{\text{maximize}}\quad & \min_{v_i\in V}\left\lfloor\frac{E_{v_i}}{e_{v_i}}\right\rfloor\IEEEyesnumber\IEEEyessubnumber*\label{eqn:opt1}\\
        \text{subject to}\quad & \sum_{(v_{i},v_{j})\in\delta(\mathcal{U})}x_{v_{i},v_{j}}\geq1,\quad& \forall\mathcal{U}\subseteq V\backslash\{v_0\},\label{eqn:opt2}\\
        & \sum_{(v_{i},v_{j})\in\delta(v_{i})}x_{v_{i},v_{j}}=1,& \forall v_i\in V\backslash\{v_0\},\label{eqn:opt3}\\
        & x_{v_{i},v_{j}}\in\brparen{0,1},& \forall v_i,v_{j}\in V,
\end{IEEEeqnarray}
where $\delta(\mathcal{S})$ is the set of edges $\brparen{(u,v):u\in S,v\notin S}$, and $x_{v_{i},v_{j}}=1$ if $v_i$ is a child of $v_j$ and $0$ otherwise. The constraint \eqref{eqn:opt2} ensures that all sensors are connected, and the constraint \eqref{eqn:opt3} ensures that each sensor can only transmit to one parent node at a time. The optimization problem in \eqref{eqn:opt} is NP-hard since it is a generalization of the NP-hard problem in \cite{Wu2008}. To approximate the complexity of the problem, we remark that if the topology is viewed as an undirected spanning tree, the number of all possible spanning trees in this network is $N^{N-2}$ by the Cayley's formula \cite{Harary1969}. Although heuristic rules can reduce the number search candidates, enumerating all potential solutions is still infeasible for a reasonable value of $N$. We propose an anytime DRL-TC algorithm that focuses on more promising areas in the search space given limited computing resources, and approaches the optimal solution with an increasing amount of computational power.

\section{The Proposed Deep Reinforcement Learning-based Topology Control Algorithm}

\subsection{Formation of Arborescence as a Markov Decision Process}
To apply reinforcement learning to the problem formulated in the previous section, we start by constructing a valid arborescence rooted at the gateway $v_0$. In each step, we select a sensor that has not been connected and connect it to a sensor or to the gateway on the tree, until all sensors are connected. This procedure can be described by a fully observable finite-horizon Markov decision process (MDP) of a $4$-tuple $\brparen{\mathcal{S},\mathcal{A},\mathcal{T},\mathcal{R}}$, as shown in Fig. \ref{fig:mdp}. At each step $t\in[0,N]$, the state of the system $s_t\in\mathcal{S}$ is the current adjacency matrix $s_t=\sqparen{x_{v_i,v_j}}_{v_i,v_j\in V}$ of the network. The action $a_t\in\mathcal{A}$ is the choice of the next sensor that will connect to the tree, or equivalently $x_{v_i,v_j}=1$ where sensor $v_i$ is to be connected to sensor (or the gateway) $v_j$ on the tree. The system then evolves to the next state $s_{t+1}$, with a deterministic transition matrix $\mathcal{T}(s,a)$ in this case. The reward at step $t$ is undetermined until the terminal state $s_N$ (i.e., all sensors are connected to the tree) is reached. Then, the objective value of \eqref{eqn:opt}, the lifetime of the WSN $r_N\in\mathcal{R}=\mathbb{R}^+$, is propagated back, as the reward for every action along the state trajectory.

\subsection{Approximating Policy and Value Functions Using a DNN}

\begin{figure}[!t]
    \centering
    \includegraphics[width=0.49\textwidth]{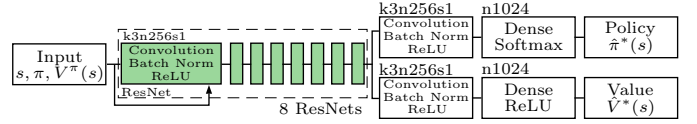}
    \caption{The architecture of the DNN in the proposed DRL-TC, which approximates the optimal policy $\hat{\pi}^*(s)$ and the optimal value function $\hat{V}^*(s)$.}
    \label{fig:dnn}
    \vspace{-1em}
\end{figure}

A stochastic policy $\pi(s)$ defines a distribution of the valid actions at a state. Under this policy, the system generates a trajectory of states and actions $h(s_t)=\brparen{s_t, a_t,\ldots,s_{N-1},a_{N-1},s_N}$, from state $s_t$ until the terminal state $s_N$. The value function $V^\pi(s)$ is defined as the expected reward of all possible trajectories, starting from state $s$ as
\begin{equation}
    V^\pi(s)\triangleq\mathsf{E}_h\sqparen{\sum_{\tau=t}^Nr_\tau|s_t=s}.
\end{equation}
We use a DNN $f_\Theta(s)$, parameterized by $\Theta$, to approximate the optimal value function $V^*(s)=\max_\pi V^\pi(s)$ together with the optimal policy $\pi^*(s)$. As shown in Fig. \ref{fig:dnn}, the input to the DNN is a training dataset $\brparen{(s,\pi(s),V^\pi(s))}$. In order to significantly increase the representational capacity of the DNN while maintaining the feasibility of training of this multiple-layer neural network, we adopt eight deep ResNet blocks proposed in \cite{He2015} on top of each other. Each ResNet consists of one convolutional layer $256$ convolutional filters each with a $3\times3$ kernel, followed by batch normalization and ReLU activation. The DNN is then split into two branches of convolutional layers followed by a dense layer with softmax and ReLU activation for the policy and the value function, respectively. The policy and value of each state predicted by the DNN $(\pi(s), V^\pi(s))=f_\Theta(s)$ contain \textit{a priori} information that guides the MCTS to search the states with high rewards and collect training datasets for the DNN in return.

Once the DNN $(V^\pi(s),\pi(s))=f_\Theta(s)$ is trained, in order to obtain an arborescence topology of the WSN, we start at the root state $s_0=\pmb{0}$, and then sequentially choose an action $a_t\sim\pi(s_t)$ from the policy predicted by the DNN and update the state $s_{t+1}=\mathcal{T}(s_t,a_t)$ until the full tree is reached. We remark that this topology construction is a stochastic process and will converge to a solution once the DNN is trained with a sufficient number of iterations.

\subsection{Collecting Training Datasets by Using MCTS}

\begin{figure}[!t]
    \centering
    \includegraphics[width=0.49\textwidth]{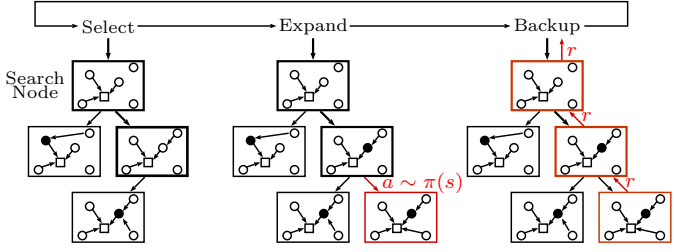}
    \caption{The procedure of the MCTS: The MCTS is expanded by the prediction from the DNN $(\pi(s),V^\pi(s))=f_\Theta(s)$, and collects training datasets in more promising search areas in return.}
    \label{fig:mcts}
    \vspace{-1em}
\end{figure}

The DNN requires a training dataset of states, policies, and values so as to fit the DNN as a function approximator. A naive approach is to enumerate and collect all states and their values as the training dataset. However, this approach will overfit the DNN and become infeasible when the state space is large. Instead of using heuristic rules to reduce the number of search candidates, we use  MCTS \cite{Browne2012} to efficiently collect training datasets in more promising areas of the search space.

\begin{table}[!t]
    \vspace{0.5em}
    \renewcommand{\arraystretch}{1.3}
    \renewcommand{\algorithmicrequire}{\textbf{Input:}}
    \renewcommand{\algorithmicensure}{\textbf{Output:}}
    \hrule\hrule
    \begin{tabular}{l}
        \hspace{-0.7em}\textbf{Algorithm 1: \textsc{Mcts}$(s)$ subroutine of the proposed DRL-TC algorithm}
    \end{tabular}
    \hrule
    \begin{algorithmic}[1]
        \REQUIRE DNN $f_\Theta(s)$; visiting counts $M(s,a)$; \textit{a priori} policy $\pi(s)$; state-action values $Q^\pi(s,a)$;
        \ENSURE Visiting counts $M$\\
        \vspace{0.5em}
        \textbf{exit conditions of recursion}\\
        \IF{$s$ is the terminal state}
            \RETURN $r$
        \ENDIF\\
        \vspace{0.5em}
        \textbf{expand to a new search leaf}\\
        \IF{$s$ has not been visited}
            \STATE $\pi(s),V(s)\leftarrow f_\Theta(s)$;
            \STATE get all valid actions for state $s$;
            \STATE re-normalize $\pi(s)$ for all valid actions;
            \STATE $M(s)\leftarrow1$;
            \RETURN $V(s)$
        \ENDIF\\
        \vspace{0.5em}
        \textbf{calculate UCBs}\\
        \STATE initialize $U\leftarrow\emptyset$;
        \FORALL{valid actions $a$}
            \STATE $U(s,a)\leftarrow Q^\pi(s,a)+c \pi(s,a)\frac{\sqrt{M(s)}}{1+M(s,a)}$;
        \ENDFOR\\
        \vspace{0.5em}
        \textbf{choose action and recursively search at the next state}\\
        \STATE $a\leftarrow\arg\max_aU(s,a)$, randomly tie-breaking;
        \STATE $s\leftarrow\mathcal{T}(s,a)$;
        \STATE recursively search at the new state $V(s)=\textsc{Mcts}(s)$;\\
        \vspace{0.5em}
        \textbf{update tree states}\\
        \STATE $Q^\pi(s,a)\leftarrow\frac{M(s,a)Q^\pi(s,a)+V(s)}{N(s,a)+1}$;
        \STATE $M(s,a)\leftarrow M(s,a)+1$
        \STATE $M(s)\leftarrow M(s)+1$;
        \RETURN $V(s)$
    \end{algorithmic}
    \hrule\hrule
    \vspace{-1em}
\end{table}

The procedure of the MCTS subroutine in DRL-TC is illustrated in Fig. \ref{fig:mcts}. Each node on the search tree represents a $5$-tuple data $(s,a,M(s,a),\pi(s),Q^\pi(s,a))$, where $s$ is the state of the WSN, $a$ is the action at the state, $M(s,a)$ is the total number of visits of $(s,a)$ on the search tree, $\pi(s)$ is a prior probability of valid actions predicted by the DNN, and $Q^\pi(s,a)$ is the state-action value, which is defined as the expected reward starting from state $s$ and taking the action $a$
\begin{equation}
    Q^\pi(s,a)\triangleq\ES{\sum_{\tau=t}^Nr_\tau|s_t=s,a_t=a}.
\end{equation}
At each search step $t<N$, the action that maximizes the upper confidence bound (UCB) \cite{Rosin2011} is selected, i.e.,
\begin{equation}\label{eqn:ucb}
    a_t=\arg\max_a\paren{Q^\pi(s,a)+c \pi(s,a)\frac{\sqrt{M(s)}}{1+M(s,a)}},
\end{equation}
where $M(s)\triangleq \sum_{b\in\mathcal{A}}M(s,b)$ is the visiting count for the state $s$ regardless of actions, and $c$ is a hyper-parameter that controls the level of exploration. Intuitively, this selection strategy initially prefers the actions with high prior probability $\pi$, but asymptotically prefers the actions with high state-action value $Q^\pi$. When the search reaches the termination state, i.e., $t=N$, a reward is obtained and propagated along the search path back to the root state for all the states visited and actions taken. The $Q^\pi$ values on the path are updated by the new average of the values on the nodes accordingly.

The details of the MCTS 
are described in \textbf{Algorithm 1}. Each search starts at a certain state and recursively searches the next state until a new leaf state or the terminal state is reached. By doing multiple MCTSs at each state, an \textit{a posteriori} visiting count $M(s)$ is collected as part of the training dataset
used to update the DNN in the next iteration.

\subsection{Self-Configuring DRL-TC Algorithm}

\begin{table}[!t]
    \vspace{1em}
    \renewcommand{\arraystretch}{1.3}
    \renewcommand{\algorithmicrequire}{\textbf{Input:}}
    \renewcommand{\algorithmicensure}{\textbf{Output:}}
    \hrule\hrule
    \begin{tabular}{l}
        \hspace{-0.7em}\textbf{Algorithm 2: The proposed DRL-TC algorithm}
    \end{tabular}
    \hrule
    \begin{algorithmic}[1]
        \REQUIRE Number of iterations $N_i$; number of episodes $N_e$; number of tree searches $N_m$; minibatch size $B$; learning rate $\alpha$;
        \ENSURE Network topology control DNN $f_\Theta(s)$ \\
        \vspace{0.5em}
        \STATE training dataset $E\leftarrow\emptyset$;
        \FOR{$i$ \textbf{from} $1$ \textbf{to} $N_i$}
            \STATE $s\leftarrow\pmb{0}$
            \FOR{$e$ \textbf{from} 1 \textbf{to} $N_e$}
                \STATE $M\leftarrow\emptyset$
                \FOR{$m$ \textbf{from} $1$ \textbf{to} $N_m$}
                    \STATE \textsc{Mcts}$(s)$
                \ENDFOR
                \STATE normalize the visiting counts $M(s)$ obtained from \textsc{Mcts}$(s)$
                \STATE $E\cup\brparen{(s,M(s),V)}$
                \IF{$s$ is the terminal state}
                    \STATE obtain the reward $r$ and update $V$ by $r$ for all $s$ in iteration $e$
                \ELSE
                    \STATE choose an action $a\sim M(s)$
                \STATE $s\leftarrow\mathcal{T}(s,a)$
                \ENDIF
            \ENDFOR
            \STATE shuffle $E$
            \STATE train DNN $f_\Theta(s)$ with a minibatch size of $B$ and learning rate of $\alpha$
        \ENDFOR
    \end{algorithmic}
    \hrule\hrule
    \vspace{-1em}
\end{table}

In short, the proposed DRL-TC alternates between the training of the DNN and MCTS, where the DNN provides an \textit{a priori} policy that guides the MCTS, and then the MCTS returns \textit{a posteriori} visiting counts and state values that are used to update the DNN. In this manner, with limited amount of computing resources, the proposed DRL-TC algorithm will focus more on promising areas to search and converges to a solution with a high reward.

The proposed DRL-TC algorithm can also adapt to the dynamics of the environment. For example, when suddenly adding or removing sensors, some actions become available or obsolete by the topology rules. In a new run of the MCTS, the policy $\pi$ returned by the DNN for the state will be re-normalized for all valid actions. Hence, the new \textit{a priori} policy $\pi(s)$ reflects the changes of the network but still correlates with the historical data. New training datasets will be collected by the MCTS and used to update the DNN. If we assume that the network change is slower than the training time which depends on the available computing resources, the proposed DRL-TC algorithm is capable of tracking the dynamics of the network and re-configuring the network topology accordingly. The complete algorithm of the proposed DRL-TC is described in \textbf{Algorithm 2}.

\section{Simulation Results and Discussions}

\subsection{Simulation Settings}
To evaluate the performance of the proposed DRL-TC algorithm, we consider a WSN with one gateway and nineteen sensors randomly scattered in a circular area with a radius of \SI{1000}{\meter}. Each sensor, with an initial energy of \SI{1}{\joule}, uniformly generates sensing data between $500$ and $1000$ bits in each round of transmission. We assume that all sensors have sufficient amount of time to transmit the data in each round. The energy dissipation per bit of processing is set to $\epsilon^\text{P}_{v_i}=$\SI{50}{\nano\joule\per\bit} for all sensors. The power amplification constant is set to $\rho=$\SI{1}{\pico\joule\per\meter\squared{\per\bit}}.

In each iteration of the algorithm, $N_e=10$ episodes of training examples are collected from the MCTS with $N_m=100$ searches at each state. The minibatch size is $B=16$ and the learning rate is $\alpha=10^{-6}$. We use the ADAM optimizer \cite{Kingma2015} to train the DNN. After each iteration of training, we evaluate the performance of the algorithm by using the DNN to construct $100$ network topologies and average the results.

\subsection{Convergence and Performance}

\begin{figure}[t]
    \setlength\abovecaptionskip{-0.3\baselineskip}
    \centering
    \includegraphics[width=0.49\textwidth]{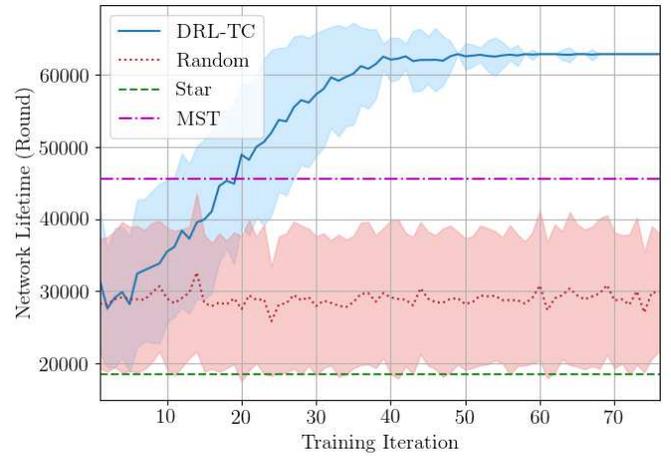}
    \caption{Convergence and performance of the proposed DRL-TC algorithm, compared with three heuristic approaches: star topology, random topology, and MST topology.}
    \label{fig:opt}
    \vspace{-1em}
\end{figure}

First, we demonstrate the convergence and performance of the proposed DRL-TC algorithm. The solid line in Fig. \ref{fig:opt} shows the average and the standard deviation (indicated by the shadowed region) of the network lifetime of $100$ realizations returned by the DNN after each training iteration. The algorithm converges after about $60$ iterations, as indicated by the diminishing standard deviation. Fig. \ref{fig:opt} also compares the performance of the proposed DRL-TC algorithm with three heuristic approaches: star topology, where all sensors connect to the gateway; random topology, where each sensor randomly chooses a node to connect to; and minimum spanning tree (MST) topology, where the MST weighted by the Euclidean distances between the nodes is formed. The star topology has the shortest network lifetime due to the high transmitting energy consumption at the edge sensors far from the gateway. The random topology shows a longer average network lifetime but with a large variance. The MST topology further improves the network lifetime by reducing the overall transmitting distance. Our proposed DRL-TC algorithm surpasses the performance of these heuristic approaches to a large extent, along with a very small variance when the algorithm converges.

\begin{figure*}[!t]
    \setlength\abovecaptionskip{-0.3\baselineskip}
    \centering
    \includegraphics[width=\textwidth]{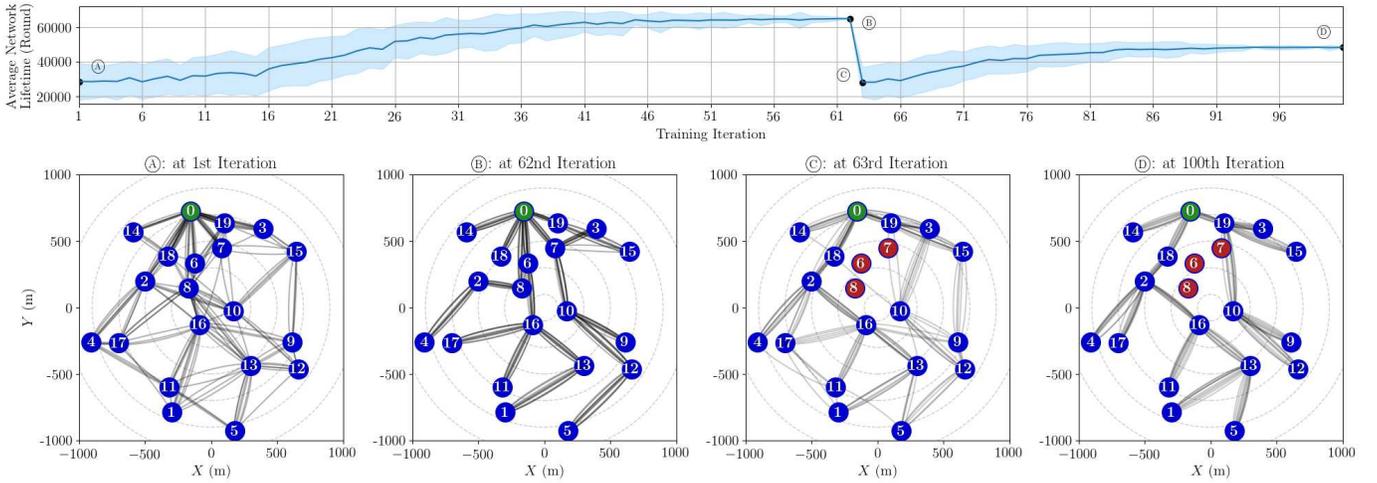}
    \caption{The evolution of the training process. Top: The DRL-TC adapts to the interruption of the sensors {\textcircled{\tiny 6}}, \textcircled{\tiny 7} and \textcircled{\tiny 8} at the $63$rd iteration and keeps improving the average network lifetime in terms of increasing its mean value and decreasing its variance (as indicated by the shadowed region). Bottom: $100$ topologies given by the DRL-TC algorithm overlaying on top of each other at the $1$st, $62$nd, $63$rd and $100$th iteration. Node \textcircled{\tiny 0} is the gateway.}
    \label{fig:evo}
    \vspace{-1em}
\end{figure*}

Figure \ref{fig:evo} demonstrates the capability of the proposed DRL-TC of adapting to sudden changes of the WSN. The top plot in Fig. \ref{fig:evo} shows the average network lifetime after each training iteration, while Figs. \ref{fig:evo}{\textcircled{\scriptsize A}} to {\textcircled{\scriptsize D}} show $100$ topologies given by the DRL-TC algorithm overlaying on top of each other after the $1$st, $62$nd, $63$rd, and $100$th iteration. As shown in Fig. \ref{fig:evo}{\textcircled{\scriptsize A}}, at the first iteration, the DRL-TC randomly explores the search space because the DNN does not have any \textit{a priori} information about the state values. After $62$ iterations, the algorithm converges to a solution with a very high confidence, as indicated by the clear paths between the sensors in Fig. \ref{fig:evo}{\textcircled{\scriptsize B}}. Then, just before the $63$rd iteration, sensors {\textcircled{\scriptsize 6}}, \textcircled{\scriptsize 7} and \textcircled{\scriptsize 8} are disabled and disconnected from the WSN, and the DRL-TC starts to re-configure the network. The new topologies, as shown in Figs. \ref{fig:evo}{\textcircled{\scriptsize C}}, are still correlated to the historical data as shown in \ref{fig:evo}{\textcircled{\scriptsize B}}. This is another advantage of the proposed DRL-TC algorithm in that it does not need to restart from scratch when the network condition changes. Eventually, the algorithm converges to another solution for the new network with a slightly smaller network lifetime due to the fact that the remaining sensors need to consume more energy to offload the data which was originally routed by the removed sensors.

\section{Conclusion and Future Work}
In this paper, we proposed a novel and unified deep reinforcement learning-based topology optimization algorithm for energy-efficient deployments of WSNs. The proposed DRL-TC algorithm is capable of adapting to the changes of the environment and shows better performance compared to other heuristic approaches to a large extent.  The framework of DRL-MCTS has a great potential in WSNs where online training is possible without intervening with the network service. In addition, with the ever-increasing computational power, we envision the emergence of other promising applications of DRL-MCTS for topology control in self-organized and fully autonomous networks of IoT in the 5G era.

\balance






\bibliographystyle{IEEEtran}
\bibliography{BSTcontrol,IEEEabrv,library}
%



\end{document}